\documentclass[aps,twocolumn]{revtex4}
\usepackage{epsfig}
\usepackage{amsmath}

\begin{document}

\title
{Thermal conductance of a two-level atom
coupled to two quantum harmonic oscillators}

\author{Pedro H. Guimar\~aes, Gabriel T. Landi and M\'ario J. de Oliveira}

\affiliation{
Instituto de F\'{\i}sica,
Universidade de S\~{a}o Paulo, \\
Rua do Mat\~ao, 1371,
05508-090 S\~{a}o Paulo, S\~{a}o Paulo, Brazil}

\date{\today}

\begin{abstract}

We have determined the thermal conductance of a system consisting of a
two-level atom coupled to two quantum harmonic oscillators in contact with heat
reservoirs at distinct temperatures. The calculation of the
heat flux as well as the atomic population and the rate of entropy production
are obtained by the use of a quantum Fokker-Planck-Kramers equation
and by a Lindblad master equation. The calculations are performed
for small values of the coupling constant. The results coming from
both approaches show that the conductance is proportional to the 
coupling constant squared and that,
at high temperatures, it is proportional to the inverse of temperature.

%05.30.-d Quantum statistical mechanics
%03.65.Yz Open (quantum) system
%05.10.Gg Stochastic analysis methods (Fokker-Planck, Langevin, etc.)
%42.50.Ct Quantum description of interaction of light and matter; related experiments
%05.60.Gg Quantum transport

\end{abstract}

\maketitle

%--------------------------------------------------------------------
\section{Introduction}

The simplest model for the interaction of atoms with radiation
is given by the Rabi model 
\cite{rabi1937,scully1997,larson2007,braak2011,larson2012,yu2012,xie2016} 
which predicts interesting properties such as the quantum Rabi oscillations
\cite{scully1997,raimond2001}.
It has a variety of applications, for instance,
in quantum optics \cite{scully1997}, 
quantum information \cite{raimond2001}, and the polaron problem \cite{holstein1959}.
The Rabi model model couples a two-level atom
to a single mode field and is represented by the quantum Hamiltonian
\begin{equation}
{\cal H} = \hbar\omega a^\dagger a + \hbar g (a^\dagger+a) \sigma_x
+ \frac{\hbar\Omega}2 \sigma_z,
\label{3}
\end{equation}
where $\Omega$ is the atomic transition frequency,
$\omega$ is the mode field frequency
and $g$ is the coupling parameter. The operators $a^\dagger$ and $a$
are the usual raising and lowering operators and we are using the notation 
$\sigma_x$, $\sigma_y$ and $\sigma_z$ for the Pauli matrices. 

The Rabi model can be extended to the case of two-mode field
by coupling the two-level atom to two harmonic oscillators. 
In this case the quantum Hamiltonian reads \cite{chilin2015}
\begin{equation}
{\cal H} = \hbar\omega (a_1^\dagger a_1 + a_2^\dagger a_2)
+ \hbar g (a_1^\dagger+a_1+a_2^\dagger+a_2) \sigma_x
+ \frac{\hbar\Omega}2 \sigma_z,
\label{2}
\end{equation}
where we are considering the two modes with the same
frequency $\omega$. The raising and lowering operators
obey the commutation relation $[a_i,a_i^\dagger]=1$.
In addition to the description of the interaction of matter with
radiation, the Hamiltonian (\ref{2}) may have other interpretations.   
For instance, in the context of superconducting circuits \cite{schoelkopf2008},
it describes the coupling
between a Josephson junction \cite{martinis2004,clarke2008,devoret2008},
represented by the two-level atom, and two transmission lines \cite{blais2004},
represented by the two harmonic oscillators.

The Hamiltonian (\ref{2}) can also be written in terms of space and
momentum variables in the equivalent form
\begin{equation}
{\cal H} = \frac{1}{2m}(p_1^2 + p_2^2) + \frac{m\omega^2}2 (x_1^2 + x_2^2)
+ \hbar\varepsilon  (x_1+x_2) \sigma_x + \frac{\hbar\Omega}2 \sigma_z,
\label{1}
\end{equation}
where $x_i$ and $p_i$ are the pair of canonical conjugated variables,
which are
related to the lowering and raising operators $a_i$ and $a_i^\dagger$ by 
$x_i = (\hbar/2m\omega)^{1/2}(a_i^\dagger+a_i)$
and $p_i = i(\hbar m \omega/2)^{1/2}(a_i^\dagger-a_i)$,
which obey the commutation relation $[x_i,p_i]=i\hbar$.
The parameter $\varepsilon$ is related to the coupling constant $g$
by $g =(\hbar/2m\omega)^{1/2}\varepsilon$.

Here we will consider the system described by the Hamiltonian (\ref{2}),
or its equivalent form (\ref{1}),
as being coupled to heat reservoirs at distinct temperatures.
More precisely one oscillator is in contact with a heat reservoir at
a higher temperature and the other oscillator is in contact with a
heat reservoir at a lower temperature. This arrangement allows us
to calculate the thermal conductance \cite{saito1996,asadian2013,tome2010,oliveira2016}
as well as the rate of the
entropy production \cite{tome2010,oliveira2016,carrillo2015,carrillo2016}
and the atomic population \cite{scully1997}, which are the main
purpose of the present study. This calculation is achieved by the use
of a quantum Fokker-Planck-Kramers (FPK) equation \cite{oliveira2016},
understood as the canonical quantization of the ordinary
FKP equation. The calculation of the conductance and the atomic
population were also performed by the use of a master equation
in the Lindblad form \cite{asadian2013,breuer2002}.
In both approaches the calculations were performed
for small values of the coupling constant. It is found that the
conductance is proportional to the square of the coupling constant and that,
at high temperatures, it is proportional to the inverse of temperature.

%--------------------------------------------------------------------
\section{Quantum FPK equation}

%------------------------------------------------
\subsection{Contact with heat reservoirs}

The contact of the system described by the Hamiltonian (\ref{1})
to heat reservoirs at temperatures $T_i$ is described by
the use of the quantum FPK equation \cite{oliveira2016}
\begin{equation}
i\hbar \frac{d\rho}{dt} = [{\cal H},\rho] - \sum_i [x_i,J_i(\rho)],
\label{5}
\end{equation}
where $\rho$ is the density matrix and
\begin{equation}
J_i(\rho) = -\frac{\gamma}{2}(\rho\, G_i + G_i^\dagger \rho)
- \frac{\gamma m}{i\hbar\beta_i}[x_i,\rho],
\label{9}
\end{equation}
and $\gamma$ is the dissipation constant, $\beta_i=1/k_B T_i$,
and $G_i$ is the operator \cite{oliveira2016},
\begin{equation}
G_i = -\frac{m}{i\hbar\beta_i}(e^{\beta_i{\cal H}} x_i e^{-\beta_i{\cal H}} -x_i).
\end{equation}
Notice that, in this description, each oscillator is in contact with one heat reservoir.
When the temperatures of the reservoirs are the same,
it follows that the equilibrium Gibbs distribution
$\rho_0=(1/Z)e^{-\beta{\cal H}}$
is the stationary solution of equation (\ref{5}) because
$J_i(\rho_0)=0$ and because $\rho_0$ commutes with ${\cal H}$.
In other words, the quantum FPK equation (\ref{5}) guarantees the
correct thermalization of the system.

From the quantum FPK equation one can determine
the time derivative of the average of energy 
$U=\langle{\cal H}\rangle={\rm Tr}\{{\cal H}\rho\}$.
It is given by
\begin{equation}
\frac{dU}{dt} =  - \sum_i \phi_i,
\end{equation}
where $\phi_i$ is the heat flux from the sytem to the reservoir~$i$,
\begin{equation}
\phi_i = \frac1{i\hbar}{\rm Tr}\{[x_i,J_i] {\cal H}\}
=-\frac1m {\rm Tr}\{p_iJ_i\}.
\label{19}
\end{equation}
From the expression of $J_i$, given by (\ref{9}), we get the heat
flux in term of averages,
\begin{equation}
\phi_i = \frac{\gamma}{2m}\langle G_ip_i + p_iG_i^\dagger\rangle
- \frac{\gamma}{\beta_i}.
\label{19a}
\end{equation}

We can also determine the time derivative of the entropy
of the system $S=-k_B{\rm Tr}\rho\ln\rho$. It is given by
\begin{equation}
\frac{dS}{dt} = \Pi - \Phi,
\end{equation}
where $\Pi$ is the rate of entropy production \cite{oliveira2016},
\begin{equation}
\Pi = \frac{k_B}{i\hbar}\sum_i {\rm Tr}\{[x_i,J_i](\ln\rho + \beta_i{\cal H})\},
\label{13}
\end{equation}
and $\Phi$ is the flux of entropy from the system to the reservoirs,
\begin{equation}
\Phi = \sum_i \frac{\phi_i}{T_i}.
\end{equation}

In the case of two reservoirs at temperatures $T_1$ and $T_2$,
and in the stationary state, $\phi_1+\phi_2=0$ and $\Pi=\Phi$,
which allow us to write the entropy production rate in the form
\begin{equation}
\Pi = \phi\left(\frac1{T_2}-\frac1{T_1}\right),
\label{25}
\end{equation}
where $\phi=-\phi_1=\phi_2$ is interpreted as the heat flux 
from reservoir 1 to the system, and equal to the heat flux
from the system to the reservoir 2.

When the temperatures are the same the expression (\ref{13})
reduces to
\begin{equation}
\Pi = \frac{k_B}{i\hbar} \sum_i
{\rm Tr} \{[x_i,J_i] (\ln\rho - \ln\rho_0)\},
\end{equation}
where $\rho_0$ is the equilibrium distribution
$\rho_0=(1/Z)e^{-\beta{\cal H}}$.

%------------------------------------------------
\subsection{Thermal Conductance}

The calculation of $G_i$ from Hamiltonian (\ref{1})
is not an easy task. However, for small values of the coupling constant
$\varepsilon$, the calculation is straightforward, although 
cumbersome, and gives
\begin{equation}
G_i = c_i p_i + ib_i x_i + \varepsilon(\mu_i\sigma_y + i\lambda_i \sigma_x),
\label{34}
\end{equation}
where 
\begin{equation}
c_i=\frac{\sinh\beta_i\hbar\omega}{\beta_i\hbar\omega},
\end{equation}
\begin{equation}
b_i=\frac{m}{\beta_i\hbar}(\cosh\beta_i\hbar\omega -1),
\end{equation}
\begin{equation}
\mu_i=\frac{\hbar\Omega}{\Lambda}
\left(- \frac{\sinh\beta_i\hbar\Omega}{\beta_i\hbar\Omega}
+ \frac{\sinh\beta_i\hbar\omega}{\beta_i\hbar\omega}\right),
\end{equation}
\begin{equation}
\lambda_i=\frac{1}{\beta_i \Lambda}\left(\cosh\beta_i\hbar\Omega 
- \cosh\beta_i\hbar\omega\right),
\end{equation}
and $\Lambda=\Omega^2 -\omega^2$.

Given $G_i$, the next task is to solve the quantum FPK equation
(\ref{5}) in the stationary state. However, instead of solving
for $\rho$, we will solve for the correlations,
such as $\langle x_i\sigma_x\rangle$, $\langle p_i\sigma_x\rangle$, 
$\langle x_i\sigma_y\rangle$, $\langle p_i\sigma_y\rangle$, and
$\langle \sigma_z\rangle$. That is,
from equation (\ref{5}) we set up evolution equations for these
quantities and solve them in the stationary regime.
%We wish to obtain the heat flux through the system, 
%in the stationary state, which we have denoted by $\phi$. The heat flux
%$\phi$ equals $-\phi_1$, where $\phi_1$ is given by expression (\ref{19}).
%From equation (\ref{19a}), it follows that the heat flux is related to the
%correlations such as $\langle p_1\sigma_y\rangle$ and $\langle p_1\sigma_x\rangle$.
%To find these correlation and others we start by setting up
%time evolution equations for the correlations. 
If $\langle A\rangle={\rm Tr}\{A\rho\}$ is one of these correlations, then
from the quantum FPK equation (\ref{5}) we reach the following formula
for the time evolution
\begin{multline}
i\hbar\frac{d}{dt}\langle A\rangle = \langle[A,{\cal H}]\rangle + \\
+ \sum_i\left(\frac{\gamma}{2}\langle G_i B_i + B_i G_i^\dagger\rangle
+ \frac{\gamma m}{i\hbar\beta_i}\langle[B_i,x_i]\rangle\right),
\label{11}
\end{multline}
where $B_i$ denotes the commutator $B_i=[A,x_i]$. 
%new material
Using equation (\ref{11}) and the expression (\ref{34}) for $G_i$,
we have obtained the evolution equations for the correlations
$\langle x_i\sigma_x\rangle$, $\langle x_i\sigma_y\rangle$, 
$\langle p_i\sigma_x\rangle$, $\langle p_i\sigma_y\rangle$, and
$\langle\sigma_z\rangle$.
In the stationary state these equations yields 
the following set of equations
\begin{equation}
\langle x_1\sigma_y\rangle + \langle x_2\sigma_y\rangle = 0,
\label{33a}
\end{equation}
\begin{equation}
m\omega^2 \langle x_i\sigma_x\rangle + \hbar\varepsilon
+ \Omega \langle p_i\sigma_y\rangle + \gamma c_i \langle p_i\sigma_x\rangle = 0,
\label{33b}
\end{equation}
\begin{equation}
\langle p_i\sigma_x\rangle - m\Omega \langle x_i\sigma_y\rangle = 0,
\label{33c}
\end{equation}
\begin{equation}
m\omega^2 \langle x_i\sigma_y\rangle - \Omega \langle p_i\sigma_x\rangle
+\gamma c_i \langle p_i\sigma_y\rangle + \gamma \varepsilon\mu_i
- \gamma \varepsilon\lambda_i\langle \sigma_z \rangle = 0,
\label{33d}
\end{equation}
\begin{equation}
\langle p_i\sigma_y\rangle + m\Omega \langle x_i\sigma_x\rangle
- 2\varepsilon \frac{u_i}{\omega^2}\langle\sigma_z\rangle = 0,
\label{33e}
\end{equation}
where 
\begin{equation}
u_i = \hbar\omega\left(\frac1{e^{\beta_i\hbar\omega}-1}+\frac12\right).
\end{equation}
%new material
Equations (\ref{33a}), (\ref{33b}), (\ref{33c}), (\ref{33d}), and (\ref{33e}),
constitute a set of nine equations because the last four 
are valid for $i=1,2$. 
It is a closed set of nine equations for the nine variables
$\langle\sigma_z\rangle$,
$\langle x_1\sigma_x\rangle$, $\langle x_2\sigma_x\rangle$, 
$\langle x_1\sigma_y\rangle$, $\langle x_2\sigma_y\rangle$,
$\langle p_1\sigma_x\rangle$, $\langle p_2\sigma_x\rangle$,
$\langle p_1\sigma_y\rangle$, $\langle p_2\sigma_y\rangle$ 
and is valid for small values of the coupling constant $\varepsilon$.
With the exception of $\langle\sigma_z\rangle$, all variables
are of the order $\varepsilon$. Up to the first order in $\varepsilon$,
on the other hand, the variable $\langle\sigma_z\rangle$ is
independent of $\varepsilon$ and is finite as we shall see.
The solution of the set of equations above is straightforward
and give all the nine quantities in closed forms.

\begin{figure}
\epsfig{file=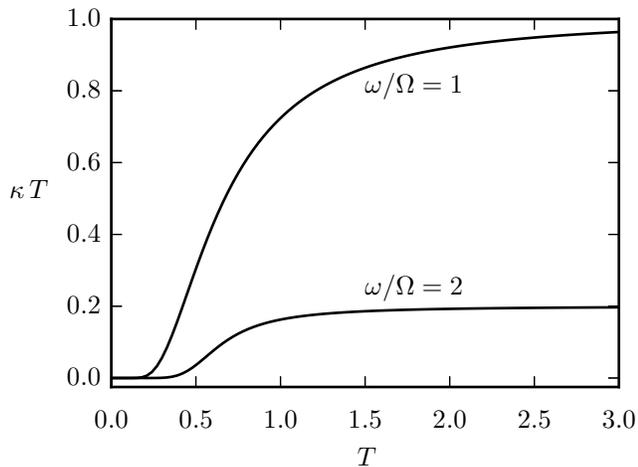}
\caption{Thermal conductance $\kappa$ times temperature $T$ as a function of $T$
according to the quantum FPK approach, equation (\ref{28}), for two values of the 
ratio $\omega/\Omega$. The parameters and constants 
$g$, $\gamma$, $\Omega$, $\hbar$ and $k_B$ are taken to be equal to unity.}
\label{fpk}
\end{figure}

To determine the heat flux, we should use equation (\ref{19a}). 
However, instead of doing so, we use 
%new material
a simpler form, obtained as follows. From equation (\ref{11}),
the evolution equation for $\langle p_i^2\rangle/2m$ is
\begin{equation}
\frac1{2m}\frac{d}{dt}\langle p_i^2\rangle
= - \frac{\omega^2}{2} \langle p_ix_i+x_ip_i\rangle
- \frac{\hbar\varepsilon}{m}\langle p_i\sigma_x\rangle - \phi_i
\end{equation}
where $\phi_i$ is given by (\ref{19a}). In the stationary state,
the left-hand side vanishes, 
$\langle p_ix_i+x_ip_i\rangle = 0$, and we are left with the 
simpler form for the heat flux
$\phi=(\hbar\varepsilon/m)\langle p_1\sigma_x\rangle$.
Using $\langle p_1\sigma_x\rangle$ obtained 
from the solution of the set of equations (\ref{33a}), 
(\ref{33b}), (\ref{33c}), (\ref{33d}), and (\ref{33e}), we find
\begin{equation}
\phi = 
\frac{2g^2 \gamma \Omega^2 \hbar\omega \,\psi_1\psi_2(v_1-v_2)}{(\Gamma_1^2+\Lambda^2)\psi_2v_2 +
(\Gamma_2^2+\Lambda^2)\psi_1v_1},
\label{26}
\end{equation}
where
\begin{equation}
\Gamma_i = \gamma\Omega c_i,
\end{equation}
\begin{equation}
v_i = \frac{\hbar\Omega}2 \coth\frac{\beta_i\hbar\Omega}{2},
\end{equation}
\begin{equation}
\psi_i =  \frac{\sinh\beta_i\hbar\Omega}{\beta_i\hbar\Omega}.
\end{equation}
We have also used the relation $g=(\hbar/2m\omega)^{1/2}\varepsilon$.
From expression (\ref{26}) we see that the entropy production rate $\Pi$,
given by equation (\ref{25}), is positive.
Indeed, if $T_1>T_2$ then $\beta_1<\beta_2$ and $v_1>v_2$ implying
$\phi>0$ and $\Pi>0$. The same conclusion follows if $T_1<T_2$.

It is worth to write down the average  $\langle\sigma_z\rangle$,
which is related to the atomic population $n$ by $n=(1+\langle\sigma_z\rangle)/2$.
It is given by
\begin{equation}
\langle\sigma_z\rangle = - \frac{\hbar\Omega}{2}
\frac{(\Gamma_1^2+\Lambda^2)\psi_2 + (\Gamma_2^2+\Lambda^2)\psi_1}
{(\Gamma_1^2+\Lambda^2)\psi_2v_2 + (\Gamma_2^2+\Lambda^2)\psi_1v_1},
\label{27}
\end{equation}
valid up to linear terms in the couplling $\varepsilon$.

To find the conductance $\kappa$ we start by writing $T_1=T+\Delta T/2$ and
$T_2=T-\Delta T/2$. The conductance is obtained from $\kappa=\phi/\Delta T$
by taking the limit $\Delta T\to0$, and is given by
\begin{equation}
\kappa = \frac{g^2 \gamma  \Omega^2 \hbar \omega}{(\gamma^2\Omega^2
c^2+\Lambda^2)T},
\label{28}
\end{equation}
where
\begin{equation}
c = \frac{\sinh\beta\hbar\omega}{\beta\hbar\omega}.
\end{equation}
%new material
When $\Omega \gg \omega$, the regime where the transitions
of the two-level atom is decoupled from the oscillators,
it turns out that the conductance is suppressed.
This can be seen in equation (\ref{28}) where $\kappa$ vanishes
in the limit $\omega/\Omega\to0$.

When $\Delta T\to 0$, the average $\langle\sigma_z\rangle$,
given by (\ref{27}), approaches the value 
\begin{equation}
\langle\sigma_z\rangle = -\tanh\frac{\beta\hbar\Omega}2,
\end{equation}
which is precisely the result we expect to obtain when
the temperatures of the reservoirs are the same, that is,
in thermodynamic equilibrium.
 
It is worth mentioning that in the high temperature limit the
conductance reduces to 
\begin{equation}
\kappa = \frac{g^2 \gamma  \Omega^2 \hbar
\omega}{(\gamma^2\Omega^2+\Lambda^2)T},
\label{28a}
\end{equation}
because in this limit, $c\to1$. On the other regime, that is,
at low temperatures, we get
\begin{equation}
\kappa = k_B\frac{4g^2}{\gamma}(\beta\hbar\omega)^3 e^{-2\beta\hbar\omega},
\end{equation}
and $\kappa$ vanishes when $T\to0$.

The thermal conductance $\kappa$ is shown in figure \ref{fpk} as a function of
temperature. We see that $\kappa T$ approaches a constant when $T\to\infty$,
in accordance with expression (\ref{28a}) which shows that,
at high temperatures, $\kappa$ is proportional to the inverse
of temperature. 

%--------------------------------------------------------------------
\section{Lindblad Master Equation}

%------------------------------------------------
\subsection{Contact with heat reservoirs}

In this approach, the master equation, which gives the time evolution
of the density matrix $\rho$, is given by \cite{asadian2013,breuer2002} 
\begin{equation}
\frac{d\rho}{dt}
= \frac{1}{i\hbar}[{\cal H},\rho] + {\cal D}_1(\rho) + {\cal D}_2(\rho),
\label{35}
\end{equation}
where, here, ${\cal H}$ is given by equation (\ref{2}),
and ${\cal D}_i(\rho)$ is the dissipator, which is
a sum of Lindblad operators,
\begin{multline}
{\cal D}_i(\rho) =  \gamma(\bar{n}_i+1)(a_i\rho a_i^\dagger -\frac12
a_i^\dagger a_i\rho -\frac12 \rho a_i^\dagger a_i) + \\
+ \gamma \bar{n}_i (a_i^\dagger\rho a_i -\frac12 a_i a_i^\dagger\rho
-\frac12\rho a_i a_i^\dagger),
\label{20}
\end{multline}
and 
\begin{equation}
\bar{n}_i = \frac{1}{e^{\beta_i\hbar\omega}-1}
\end{equation}
is the Bose-Einstein distribution.
We remark that ${\cal D}_i$ are local Lindblad dissipators and
that equation (\ref{20}) gives an adequate description of the system
only in the regime of weak interactions, which is precisely the 
regime considered here.

From the Lindblad master equation, one can calculate again the time 
evolution of the average energy
$U=\langle\cal{H}\rangle=\rm{Tr}\{\cal{H}\rho\}$ given by equation $(7)$,
where the heat flux $\phi_i$ now reads
\begin{equation}
\phi_i = - \mathrm{Tr}\left\{\mathcal{D}_i\mathcal{H}\right\}.
\label{21}
\end{equation}

To find an expression for the entropy production rate we start
by postulating that the flux of entropy related to each
reservoir is the heat flux divided by its temperature.
Therefore the total entropy heat flux is
\begin{equation}
\Phi = \sum_i \frac{\phi_i}{T_i}.
\label{42}
\end{equation}
Next we determine the time derivative of the entropy of the
system $S=-k_B{\rm Tr}\rho\ln\rho$ and use the expression
$dS/dt=\Pi-\Phi$ to find the rate of entropy production,
\begin{equation}
\Pi = - k_B \sum_i {\rm Tr} \{{\cal D}_i (\ln\rho + \beta_i \mathcal{H})\}.
\label{39}
\end{equation}
In the stationary state, $\Pi=\Phi$, and we may use expression
(\ref{42}) to write
\begin{equation}
\Phi = \phi\left(\frac{1}{T_2}- \frac{1}{T_1} \right),
\label{42a}
\end{equation}
where $\phi=-\phi_1=\phi_2$. 

When the temperatures are the same, expression (\ref{39})
reduces to 
\begin{equation}
%\Pi = - k_B \sum_i {\rm Tr} \{{\cal D}_i (\ln\rho + \beta \mathcal{H})\},
\Pi = - k_B \sum_i {\rm Tr} \{{\cal D}_i (\ln\rho - \ln\rho_0)\},
\label{41}
\end{equation}
where $\rho_0=(1/Z)e^{-\beta{\cal H}}$ is the equilibrium 
Gibbs distribution. Equation (\ref{41}) gives the usual expression for the
rate of entropy production associated to master equation of
the Lindblad type \cite{breuer2002}.

The establishment of quations (\ref{42}) and (\ref{39})
for the heat flux and the entropy production rate requires
that the equilibrium solution of the master equation is the Gibbs state
$\rho_0=(1/Z)e^{-\beta{\cal H}}$, otherwise the entropy
production rate will not vanish in equilibrium.
This requirement means to say that ${\cal D}_i(\rho_0)$
should vanish when the temperatures are the same.
However, the local Lindblad dissipators of the type (\ref{20})
do not hold this property and should thus be understood as approximations,
valid, in the present case, in the regime of weak interactions.
The calculations of the heat flux and the entropy flux by
equations (\ref{21}) and (\ref{42}) should therefore be understood
as approximate results, but this is only because the local
dissipators (\ref{20}) are approximations.

\begin{figure}
\epsfig{file=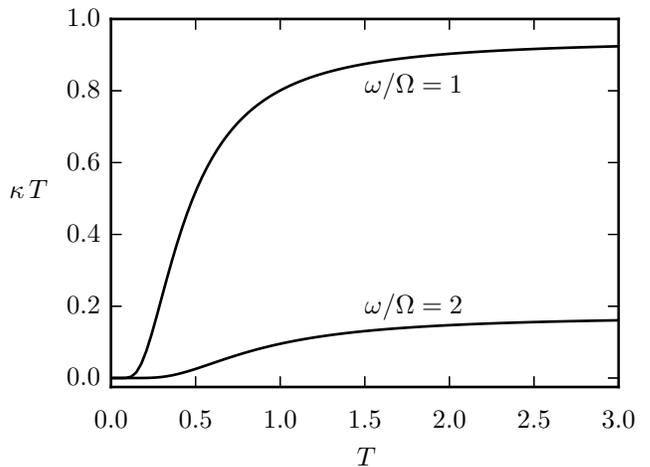}
\caption{Thermal conductance $\kappa$ times temperature $T$ as a function of $T$
according to the Lindblad master equation approach, equation (\ref{40}),
for two values of the ratio $\omega/\Omega$. The parameters and constants 
$g$, $\gamma$, $\Omega$, $\hbar$ and $k_B$ are taken to be equal to unity.}
\label{lindblad}
\end{figure}

%------------------------------------------------
\subsection{Thermal Conductance}

From the master equation we can again determine the time evolution of
the quantities $\langle a_i\sigma_x\rangle$,
$\langle a_i\sigma_y\rangle$, $\langle a_i^\dagger\sigma_x\rangle$,
$\langle a_i^\dagger\sigma_y\rangle$ and $\langle\sigma_z\rangle$. 
If $\langle A\rangle={\rm Tr}\{A\rho\}$ is one of these quantities, then from 
the master Lindblad equation we may obtain the following formula
for its time evolution
\begin{equation}
\frac{d}{dt}\langle A\rangle = \frac{1}{i\hbar}\langle[A,{\cal H}]\rangle
+ \sum_i {\rm Tr}\{A{\cal D}_i\}.
\label{51}
\end{equation}
Using these equation and the definition (\ref{20}) of ${\cal D}_i$,
we obtained the evolution of the quantities of interest.
In the stationary state, these equations give the following set of equations
\begin{equation}
\langle a_{1}\sigma_{y}\rangle +\langle a_{1}^{\dagger}\sigma_{y}\rangle=
\langle a_{2}\sigma_{y}\rangle+\langle a_{2}^{\dagger}\sigma_{y}\rangle,
\label{52a}
\end{equation}
\begin{equation}
\alpha\langle a_{i}\sigma_{x}\rangle+\Omega\langle a_{i}\sigma_{y}\rangle+ig = 0,
\label{52b}
\end{equation}
\begin{equation}
\alpha^{*}\langle a_{i}^{\dagger}\sigma_{x}\rangle
+\Omega\langle a_{i}^{\dagger}\sigma_{y}\rangle-ig = 0,
\label{52c}
\end{equation}
\begin{equation}
\alpha\langle a_{i}\sigma_{y}\rangle-\Omega
\langle a_{i}\sigma_{x}\rangle + g\left(2\bar{n}
+1\right)\langle\sigma_{z}\rangle = 0,
\label{52d}
\end{equation}
\begin{equation}
\alpha^{*}\langle a_{i}^{\dagger}\sigma_{y}\rangle
-\Omega\langle a_{i}^{\dagger}\sigma_{x}\rangle
+g\left(2\bar{n}+1\right)\langle\sigma_{z}\rangle = 0,
\label{52e}
\end{equation}
where $\alpha = \gamma/2 + i\omega$.
This is a closed set of nine equations in the
variables  $\langle\sigma_z\rangle$,
$\langle a_1\sigma_x\rangle$, $\langle a_2\sigma_x\rangle$,
$\langle a_1\sigma_y\rangle$, $\langle a_2\sigma_y\rangle$,
$\langle a_1^\dagger\sigma_x\rangle$, $\langle a_2^\dagger\sigma_x\rangle$,
$\langle a_1^\dagger\sigma_y\rangle$, and $\langle a_2^\dagger\sigma_y\rangle$.
Notice the last four equation are valid for $i=1,2$.
The set of equation is valid for small values of the coupling constant $g$
and is easily solved to get the correlations in closed forms.
Notice that, with the exception of the variable $\langle\sigma_z\rangle$,
all variables are proportional to the coupling constant $g$ whereas
$\langle\sigma_z\rangle$ does not depend on $g$.

To obtain an expression for the heat flux, we replace
the definition of ${\cal D}_i$, given by equation (\ref{20}),  
into equation (\ref{21}),
\begin{equation}
\phi_i = -\hbar\omega{\rm Tr}\{a_i^\dagger a_i {\cal D}_i\}
+\frac{\hbar g\gamma}{2}\langle(a_i^\dagger+a_i)\sigma_x\rangle
\end{equation}
Now, the evolution equation for the quantity $\langle a_i^\dagger a_i\rangle$ is
\begin{equation}
\frac{d}{dt}\langle a_i^\dagger a_i\rangle = 
-i g \langle(a_i^\dagger - a_i)\sigma_x\rangle
+ {\rm Tr}\{a_i^\dagger a_i {\cal D}_i\}
\end{equation}
In the stationary state, the left-hand side vanishes.
Summing up these two last equations, we obtain the  
the heat flux $\phi=-\phi_1$ in the form
\begin{equation}
\phi=i\hbar\omega g \langle(a_1^{\dagger}-a_1)\sigma_x\rangle
-\frac{\hbar g\gamma}{2} \langle (a_1^{\dagger}+ a_1)\sigma_x\rangle.
\end{equation}

From $\langle a_1\sigma_x\rangle$ and $\langle a_1^\dagger\sigma_x\rangle$,
obtained from the stationary solution, we get the heat flux
\begin{equation}
\phi=\frac{32g^{2}\gamma\Omega^{2}\hbar\omega}{\gamma^4+8 \gamma^2
(\omega ^2+\Omega ^2)+16\Lambda^2}\frac{\bar{n}_1-\bar{n}_2}{1+\bar{n}_1+\bar{n}_2}.
\end{equation}
As we did in the previous approach, it is worth to write down the mean atomic
population
\begin{equation}
\langle\sigma_z\rangle
= -\frac{8\omega\Omega}{\gamma^2+4 (\omega^2+\Omega ^2)}
\frac{1}{1+\bar{n}_1+\bar{n}_2}.
\end{equation}

By writing again the temperatures as $T_1=T+\Delta T/2$ and
$T_2=T-\Delta T/2$, it is possible to obtain the thermal conductance $\kappa$.
In this case it reads
\begin{equation}
\kappa=\frac{16g^{2}\gamma\Omega^{2}\hbar\omega}{\gamma^{4}+8\gamma^{2}
(\omega^{2}+\Omega^{2})+16\Lambda^{2}}\frac{1}{cT}.
\label{40}
\end{equation}
In this case, the quantity $\langle\sigma_z\rangle$ becomes
\begin{equation}
\langle\sigma_z\rangle
= -\frac{8\omega\Omega}{\gamma ^2+4 \left(\omega ^2+\Omega
^2\right)}\tanh\frac{\beta\hbar\omega}{2},
\label{29}
\end{equation}
which, as we can see, does not correspond to the correct result for the 
thermodynamic equilibrium. This incorrect result is to be expected because
local phenomenological Lindblad master equations do not lead to proper thermalization, that is,
the Gibbs probability distribution $\rho_0=(1/Z)e^{-\beta{\cal H}}$
is not the stationary solution of the Lindblad master equation (\ref{35}).

In the high temperature limit, the thermal conductance reduces to
\begin{equation}
\kappa = \frac{16g^{2}\gamma\Omega^{2}\hbar\omega}
{\gamma^{4} + 8\gamma^{2}(\omega^{2}+\Omega^{2})+16\Lambda^{2}}\frac{1}{T},
\label{28b}
\end{equation}
because $c\to1$ in this limit. At low temperatures, on the other hand,
the thermal conductance becomes
\begin{equation}
\kappa=k_B\frac{32g^{2}\gamma\Omega^{2}(\beta\hbar\omega)^{2}}
{\gamma^{4} + 8\gamma^{2}\left(\omega^{2}+\Omega^{2}\right)+16\Lambda^{2}}
e^{-\beta\hbar\omega},
\end{equation}
and $\kappa$ vanishes when $T\to0$.

The thermal conductance $\kappa$ is shown in figure \ref{lindblad} as a function of
temperature. We see that $\kappa T$ approaches a constant when $T\to\infty$,
in accordance with expression (\ref{28b}) which shows that,
at high temperatures, $\kappa$ is proportional to the inverse
of temperature. In this respect, the two approaches give similar
results.

%--------------------------------------------------------------------
\section{Discussion}

We have determined the quantum conductance of a system consisting
of two level atom coupled to two harmonic oscillators by the use
of two distinct methods. 
% material added
These two methods should be understood as two distinct theories
about open quantum systems which differ in the way the contact
of the system with the heat bath is treated. In the case of the quantum FPK,
the contact with a heat bath is advanced in terms of
a dissipation-fluctuation approach, described by each term
$[x_i,J_i]$ on the summation on the right-hand side of equation (\ref{5}).
In both cases we have set up equations
for the correlations which were determined in closed form. From these
correlations we have obtained the heat flux and the conductance,
which is found to be proportional to coupling constant squared.
At high temperature the conductance were found to be proportional
to the inverse of temperature, $\kappa\sim T^{-1}$ for both cases.
We point out that, at resonance, $\omega=\Omega$,
and assuming $\gamma$ much smaller that $\omega$,
both approaches yields the same result
$\kappa=g^2\hbar\omega/\gamma T$.
At low temperature the conductance vanishes with temperature
as $\kappa\sim e^{-k/T}/T^n$ with $n=3$ for the first method
and $n=2$ for the second method. All calculations were performed
for small values of the coupling constant.

We have also determined the atomic population, $n=(1+\langle\sigma_z\rangle)/2$.
Up to linear order in the coupling constant, it is
finite and, when the difference in the temperatures of the reservoirs
vanish, it should be identical to the equilibrium value. Indeed
this is what happens when we use the quantum FPK approach, as can be
seen in equation (\ref{27}). However, this is not the case when
we use the phenomenological Lindblad master equation with local dissipators
(\ref{35}). In this case, the atomic
population $\langle\sigma_z\rangle$, given by (\ref{29}),
differs from the equilibrium value. 

We have also determined the rate of entropy production for the case
of the quantum FPK approach. In this approach, the rate of entropy
production, according to reference \cite{oliveira2016},
is defined by equation (\ref{13}). Using this definition, we obtain
equation (\ref{25}), which shows that the rate of entropy production
is a product of the heat flux $\phi$ and
the thermodynamic force $(T_2^{-1}-T_1^{-1})$, and proven to be positive
as expected. For small values of $\Delta T=T_1-T_2$, we may write
$\Pi=\kappa(\Delta T/T)^2$ which is clearly positive because $\kappa$
is positive.

%--------------------------------------------------------------------
\section*{Acknowledgement}

G. T. L. would like to acknowledge the S\~ao Paulo Research Foundation
under grant number 2016/08721-7. P. H. G. would like to acknowledge 
the fellowship from the Brazilian agency CNPq.

%--------------------------------------------------------------------

\end{document}